# Navigating through Educational Pathways to Political Participation: A Multi-theoretical Exploration of Voting Behaviors

1. *Muhammad Hassan Bin Afzal
   Visiting Assistant Professor
   Department of Political Science and Public Service
   University of Tennessee at Chattanooga
   Hassan-Afzal@utc.edu

2. Paula Daniela Ganga
   Duke Kunshan University
   pdg23@georgetown.edu

3. Oindrila Roy
   Associate Professor of International Relations
   Cottey College
   oroy@cottey.edu

4. Kristina Thompson
   Undergraduate Research Fellow
   University of Tennessee at Chattanooga
   bqb899@mocs.utc.edu

**Abstract:**

    We investigate the determinants of voting behavior by focusing on the direct effect of educational attainment, sociodemographic characteristics, partisan identity, and political ideology on the intention to vote, registration, and turnout. We use the cumulative CCES dataset to explore voting behavior for the 2014 and 2018 midterm elections and the 2016 and 2020 general elections. We propose a new Voting Engagement Index (VEI) to assess these factors' cumulative impact on electoral participation. Our analysis shows that education consistently motivates voting behavior, while gender, race, and ethnicity significantly shape engagement levels. Mainly, Black and Middle Eastern Americans exhibit higher voting engagement, whereas Native Americans and females display lower odds of voting engagement. Although Native Americans and women express a clear intention to vote in upcoming elections with increased attainment, the intention is not fully realized in voter registration and voting during midterms and general elections. Income and home ownership also become apparent as strong predictors of voter engagement. This research contributes to understanding the changing aspects of voter motivation and participation, with implications for grassroots-level mobilization, including unheard voting voices in U.S. elections, more inclusive and just voting policies and future electoral studies.

**Keywords:** Voting Behavior, Educational Attainment, Home Ownership, Native Americans, Voting Engagement Index (VEI), Electoral participation.

**Introduction:**

In the ever-evolving Western democracy, the ongoing transition from democratic government towards modern governance (Bang & Eva, 1999) warrants revisiting and reevaluating the driving factors that ensure civic agency (Boyte, 2005; Franck, 1992) and citizen engagement (Anderson & Dodd, 2009; Smith, 2009). Exploring the profound roles of deeply ingrained multifaceted personal, social, and voting regulations helps us to understand how general populations perceive the impact of voting in their civic engagement in democracy (Alex-Assensoh, 2005). Intention to vote, access to voting registration, and voting play a role in uniting the voices of the people of a nation. With the upcoming general election 2024 in the United States, there is a spotlight on the significance of citizen engagement in voting, prompting a look at the factors that impact voter participation.

The 2020 general election recently saw a record-high turnout, driven by intense political engagement and the expansion of mail-in voting due to the COVID-19 pandemic (Stachofsky et al., 2023). The voter turnout rate stood at 61.4% during the 2016 Presidential Election, similar to the 2012 turnout of 61.8%, showcasing the effects of age groups, race and ethnicity, and educational attainment underscore voter engagement's diverse nature (Bureau, 2023b). During the 2020 General Election, voter turnout surged to 66.7%, marking the highest participation rate for over a century, which was facilitated through expanded access to voting options, such as mail-in ballots and early voting options, during the pandemic fueled this increase (Davenport, 2023; Foreman et al., 2021; Rickenbach et al., 2023; Varona, 2023). In contrast, midterm elections, such as those in 2014 and 2018, typically experience lower turnout, highlighting the varying levels of engagement between presidential and non-presidential election years (Autor et al., 2020; Bureau, 2023a; Hicks et al., 2015; National Geographic Society, 2023; Niederdeppe et al., 2021; Shaw & Petrocik, 2020). These trends underscore the importance of examining the determinants of voter participation, including education, income, and the ease of voting. This study, which utilizes data from the Cooperative Congressional Election Study (CCES) dataset, aims to examine how educational attainment, demographic elements, and voting behaviors intersected during elections like 2012 and 2016 compared to 2020, well as during midterm elections in 2014 and 2018 compared to those in 2022.

     Despite ever-evolving and expanding in-depth research, starting from the role of education as a function of self-governance (Dewey, 1916) to Partisan Mobilization theory (Dalton, 2004) on the connection between education and voting tendencies in midterm (Plane & Gershtenson, 2004) and general elections (Adams & Merrill, 2003; Basinger & Lavine, 2005; Burden, 2009; Kernell, 2014; Kogan et al., 2018; Levitin & Miller, 2013; Sosnaud et al., 2013), there still exists a nuanced gap to grasp this intricate relationship fully. Therefore, we aim to investigate this gap by analyzing the full spectrum of voting participation, starting from intention to vote, then verified registration, and finally verified voting in general and midterm elections. Hence, there is a call for an approach that considers multiple theories to thoroughly explore how education impacts voting behavior while considering the diverse elements influencing an individual's decision to participate in elections. Our research investigates the multifaceted relationship between education and voting tendencies, asking questions such as how educational attainment influences voter registration and election participation rates and how psychological

factors and political awareness fostered by experiences influence one's participation. To address this gap, we introduce the Voting Engagement Index (VEI), a metric designed to evaluate voting behavior from multiple perspectives. The VEI is an additive index that acknowledges each positive intention to vote incrementally, combines it with verified registration status, and verifies voting after elections. We aim to evaluate the VEI's impact on voting behavior to other indicators. This study intends to give ideas on tailoring efforts to increase engagement and eliminate hurdles that prevent individuals from participating in voting.

## **Motivations for the Current Study:**

The primary goal of this study is to analyze what factors motivate the potential voting-age population in the United States to express their clear intention to vote in an upcoming election. Is educational attainment sufficient for them to express their clear intention to vote and follow through with their intention through the rational next steps of registering to vote and voting in an upcoming election? We use the cumulative CCES dataset to explore the last two midterm elections and general election voting behavior among the voting-age populations. What factors drive individuals' explicit intention to vote, follow through with their intention with registration, and finally, the actual voting process? Does gender play a significant role in determining the clear expression of an intention to vote, combined with the level of attained education, which helps them progress towards registration and actual voting? We also explore the role of race and ethnicity in voting behavior among the voting-age population in the United States: does education, socioeconomic status, and partisan identity drive human beings to follow through on all three voting behaviors, such as intention to vote, registration, and final voting? We aim to explore whether intersectional sociodemographic characteristics play any role in enhancing that intention, registration, and voting or creating more perceived or actual barriers in their voting engagement.

**Literature Review**

The importance of access to education for citizens to enhance political participation has long been recognized from the start of the 20th century. Dewey (1916) emphasizes the importance of education in promoting active citizenship and political participation:

> "A society which makes provision for participation in its good of all its members on equal terms and which secures flexible readjustment of its institutions through interaction of the different forms of associated life is in so far democratic. Such a society must have a type of education which gives individuals a personal interest in social relationships and control, and the habits of mind which secure social changes without introducing disorder." (p. 116)

Here, Dewey (1916) strongly contends that education enhances citizens' active citizenship, which instills citizens' internal drive to be more active in political participation. This viewpoint is shared by historical leaders such as Noah Webster, who advocated that education should be a top priority for legislators to ensure enhanced political participation among the population (Gutmann, 1999; Webster, 1790). The profound role of education in providing sociopolitical tools for the future generation in political participation, Gutmann (1999) contends, "Democratic education is therefore a political as well as an educational ideal." Nie et al. (1996) found a substantial and constant link between formal education and political engagement, highlighting the importance of education in developing civic traits. Additionally, their research emphasizes the consensus that formal education is a significant predictor of political engagement, implying that education directly impacts civic participation (Nie et al., 1996). According to the "Standard Model" (Brady et al., 1995; Carpini & Keeter, 1996), education drives political activity. The "Preadult Socialization Model" views schooling as a key factor influencing persons' democratic engagement from a young age (Gutmann, 1999).

Another school of thought contends that political engagement among the population is influenced by combined factors that outline their social position rather than solely education (Brady et al., 2015; Jennings et al., 2009). The main argument is that education is insufficient to drive political engagement among the population; education combined with higher socioeconomic conditions is more likely to result in higher political participation. The Civic Voluntarism Model (CVM) suggests that having resources, such as education, may increase political participation and mobilization; on the other hand, the Resource Model of Political Involvement suggests that having specific resources, such as education, might increase involvement (Brady et al., 1995, 2015; Schlozman et al., 2018).

The theory of Political Socialization contends that education and social institutions, direct and indirect agents of socialization, strongly influence civic engagement, voting behavior, and political participation (Hyman, 1959). Additionally, the Partisan Mobilization theory (Dalton, 2004) recognizes the importance of party identification as an essential factor in political participation, signifying that cumulative educational experiences are more likely to influence individuals' alignment with political parties and participation in the political process.

On the other hand, several research studies strongly caution against accepting only the direct causal link between attained education and political participation (Berinsky & Lenz, 2011; Brade et al., 2016; Kam & Palmer, 2008). The importance of assessing the multidimensional effects of education on political participation is investigated through the intersectional lens of capturing education enhances civic knowledge (Condon, 2015; Persson, 2015; Quinn et al., 2013) and self-importance in voting behavior for minorities and historically overlooked community members (Caffrey-Maffei, 2019; Cichocka et al., 2024).

In existing literature and studies, researchers have explored various dimensions of voting behavior among the eligible population in the United States. These investigations separately focus on the intention to vote (Beyer et al., 2014; Bol et al., 2021; Geldrop, 2020; Glynn et al., 2009; Johnson et al., 2014; Morar & Chuchu, 2015; Soroka et al., 2009), voter registration, and actual voting (Ansolabehere & Konisky, 2006; Garnett, 2022; Minnis & Shah, 2020; Nickerson, 2015) in general and midterm elections. Additionally, self-reported turnout has been a critical area of interest. Our study aims to examine the more pronounced effect of the voting behavior spectrum that respondents follow—from their initial intention to vote through registration and ultimately to cast a validated vote in either a midterm or general election. Specifically, we seek to understand the direct impact of education across all three levels of participation: intention, registration, and voting. To quantify this engagement, we create an additive index that considers intention, validated registration, and validated voting behavior. Respondents receive a positive partial score based on their intention to vote and information on validated voter registration and actual voting in either midterm or general elections.

**Data and Methods:**

Our study utilizes the Cooperative Congressional Election Study (CCES) Cumulative dataset to examine the intricate relationship between education and voting behavior across several election cycles in the United States. We primarily focus on general elections in 2012 and 2016, with 2020 as the reference category, and midterm elections in 2014 and 2018, with 2022 as the reference category; our analysis aims to shed light on the nuanced dynamics of electoral participation over time. The dependent variables central to our study include intention to vote (pre-election wave), registration status, actual voting (voted bin), and the Voting Engagement Index (VEI).

We use a novel VEI metric developed for this research, which encapsulates the multifaceted nature of voting behavior by integrating respondents' intention to vote, verified registration status, and validated voting into a comprehensive measure of electoral engagement. The index is constructed by recoding intent_to_vote (0 = 0, 1 = 0.25, 2 = 0.5, 3 = 1)[1], reg_status_bin (0 = 0, 1 = 1), and voted_bin (0 = 0, 1 = 1), in so doing capturing incremental steps towards a clearly defined intention to vote, beginning with the 2012 general election and extending through subsequent general and midterm elections. Our empirical strategy unfolds through a series of ordinal and binomial logistic regression models, each tailored to explore different facets of the education-voting link. Model 1, the most restrictive, examines the bivariate relationship between education and each dependent variable, laying the groundwork for understanding the foundational impact of educational attainment on voting behavior. Model 2 introduces control variables[2] and concentrates on midterm elections, with 2022 as the reference category, to assess how education influences voting behavior, controlling various demographic and socio-political factors. Model 3, the most comprehensive, encompasses both midterm and general elections, controlling for the 2020 general election and the 2022 midterm as reference categories to provide a holistic analysis of the role of education in shaping voting behavior across different electoral contexts.

---

[1] The goal of the index is to recognize each non-negative intention to vote in upcoming election.
[2] In all models, we control for a range of variables including age, gender, race, income, home status, political ideology, party affiliation (Democrats, Republicans, Independents, and Other/Not Sure as the reference category), as well as participation in the general elections of 2016, 2012 (with 2020 as the reference category), and midterm elections of 2018, 2014 (with 2022 as the reference category).

To ensure consistency in our analysis, we meticulously verify that all models had consistent, complete cases, amounting to 294,999 responses for each dependent and independent variable. This careful attention to data integrity allows for robust and reliable comparisons across our models. We adopt a Last-In-First-Out (LIFO) approach, utilizing the most recent elections (2020 general election and 2022 midterm) as reference categories. We rationalize our reference category choice to probe into the specific factors that influence voting behaviors in the backdrop of expanded access during the COVID-19 pandemic and the unique circumstances surrounding the 2020 and 2022 elections. By examining the multifaceted relationship between education and voting tendencies, our study seeks to contribute to a deeper understanding of electoral participation mechanisms and how education shapes the democratic process. The descriptive statistics in Table 1 describe the variables in detail (See table 1).

**Results and Findings:**

The first set of three models explores (See Table 2) the effect of increasing the attained level of education and respondents' intention to vote in the upcoming election. The DV asks explicitly for the intention to vote, which ranges from a clear no to undecided to probably to a clear yes. We also use an "if missing case" in Stata to ensure every model has the same numbers of N= 294,999, and The first bivariate model finds that the most fundamental of our analyses revealed a significant positive association between education level and the intention to vote, with an odds ratio of 1.717 ($p < 0.001$). In simpler terms, a one-unit increase in attained education level corresponds to a 71.1% increase in the odds of moving to a more positive intention vote in the upcoming election, which is statistically significant. The bivariate coefficient signifies that at least two units of attained education ensure a clear intention to vote in the forthcoming election.

In the three models for (DV = Intention to Vote), we find that having a level of education consistently increases the likelihood of intending to vote. This trend was evident across all three models we examined. Specifically, educated individuals had odds of planning to vote, as shown by the odds ratios of 1.717 (M1), 1.669 (M2), and 1.71 (M3), and these coefficients are statistically significant. Likewise, our analysis indicates that older individuals were more inclined to express an intention to vote based on the odds ratios of 1.033 (M2) and 1.034 (M3), and these coefficients are statistically significant. On the other hand, gender seemed to impact voting intentions in these models, with females showing lower odds of planning to vote compared to males; odds ratios of 0.785 (M2) and 0.776 (M3), and these coefficients are statistically significant.

We also observe how race influenced voting intentions significantly; Black, Hispanic, and Middle Eastern individuals showed an increased odds ratio likelihood of intending to vote than their White counterparts in Models 2 and 3. The surprising factor we observe is that Native Americans show a decreased likelihood of expressing their intention to vote in upcoming elections in both models 2 and 3. Additionally, income levels and home ownership status were positively linked with higher odds of voting intention, and our analysis underscored the role of these factors. Political party affiliation indicates voting intentions, as Democrats, Republicans, and Independents all likely plan to vote compared to those without party affiliation. There is a steady decline in the intention to vote among respondents in Model 3, where the coefficients for the 2012 general election (2.6), 2014 midterm (1.2), 2016 general election (1.8), and 2018 midterm (0.86) and all these coefficients are statistically significant.

The second series of models (see Table 3) delves into the relationship between attained education levels and respondents' registration status as active voters. Consistent with our previous analyses, we maintain the same number of observations (N = 294,999) across all models to ensure comparability. The dependent variable here ensures that the participants have validated active registration status to vote in the upcoming election. In the initial bivariate model, we find an increased level of education and active voter registration status, with an odds ratio of 1.143 ($p < 0.001$). The bivariate education attainment coefficient shows that a one-unit increase in attained education corresponds to a 14.3% increase in the odds of being actively registered to vote.

We use the same set of variables for both Model 2 and Model 2 that we used for intention-to-vote models. Here, we find that, specifically in Models 2 and 3, the education trend positively influencing active voter registration persists. Specifically, the odds ratios for education level across the models are 1.143 (M1), 1.166 (M2), and 1.238 (M3), indicating that higher levels of education are consistently associated with increased odds of active voter registration, with all coefficients being statistically significant. We also observe that similar to intention to vote models, age emerges as a significant predictor, with older individuals more likely to be actively registered to vote, as indicated by the odds ratios of 1.019 (M2) and 1.024 (M3). Most interestingly, unlike the intention to vote, females exhibit a slightly higher positive effect of 1.047 (M2) and 1.039 (M3), suggesting that females are more likely than males to be actively registered to vote. Both female coefficients are statistically significant in both M2 and M3 models.

We also find that race plays a significant role in determining validated registration status among survey participants. Black individuals show a significantly increased likelihood of being actively registered at 1.591 (M2) and 1.766 ( M3), and these coefficients are statistically significant. Similar to intention-to-vote models, we observe that Native American individuals demonstrate a decreased likelihood of active registration 0.661 (M2), but the effect is not statistically significant in model 3. Socioeconomic factors such as income levels and home ownership status are positively associated with active voter registration, highlighting the influence of these variables. Political party affiliation also plays a crucial role, with Democrats, Republicans, and Independents all showing higher odds of being actively registered than those without party affiliation.

We progressed to the series of verified votes as DV regression models. We conduct a series of three binomial logistic regressions investigating the relationship between attained education levels and respondents' verified voting status in recent elections. As with our previous models, we maintain a consistent number of observations (N = 294,999) to ensure the reliability of our findings. The dependent variable in this series focuses on whether participants have a validated voting record for the election in question. In the initial bivariate model of this series, we observe a positive association between education level and verified voting, with an odds ratio of 1.236 ($p < 0.001$). The bivariate coefficient from Model 1 indicates that a one-unit increase in attained education level corresponds to a 23.6% increase in the odds of having a validated vote, highlighting the significant role of education in electoral participation, starting from intention to vote (see Table 2), verified registration (see Table 3), and finally voting (see Table 4).

As we progress with our further analysis of Models 2 and 3, the positive influence of education on verified voting remains consistent. Specifically, the odds ratios for education level are 1.236 (M1), 1.240 (M2), and 1.316 (M3), suggesting that higher levels of education are associated with increased odds of verified voting, with all coefficients being statistically significant. This upward trend is consistent with the earlier intention to vote and registration status models, and all these coefficients are statistically significant.

Similar to the last two models (Intention to vote and registration), age also emerges as a significant factor, with older individuals more likely to have a validated vote, as evidenced by the odds ratios of 1.025 (M2) and 1.031 (M3). Gender, on the other hand, appears to have a minimal impact on verified voting in these models, females with odds ratios close to 1. Similarly, race plays a statistically significant role in determining verified voting status. While Black individuals show an increased likelihood of having a validated vote in Models 2 and 3, Native American individuals demonstrate a decreased likelihood in both models, with the effect being statistically significant.

We also observe that socioeconomic factors, such as income levels and home ownership status, are positively associated with verified voting, indicating that higher income and home ownership are linked to higher odds of having a validated vote. Political party affiliation also emerges as a crucial predictor, with Democrats, Republicans, and Independents all showing higher odds of verified voting than those without party affiliation.

Now, we propose to create a Voting Engagement Index (VEI) that combines intention to vote, voter registration and verified voted status, where we investigate the relationship between attained education levels and respondents' overall voting engagement, as measured by the Voting Engagement Index (VEI). This composite index encompasses the intention to vote, validated registration, and verified voting, offering a holistic view of electoral participation. As with our previous analyses, we ensure a consistent sample size of N = 294,999 across all models.

The initial bivariate model reveals a significant positive association between education level and the VEI, with an odds ratio of 1.315 ($p < 0.001$). The bivariate coefficient indicates that a one-unit increase in attained education level corresponds to a 31.5% increase in the odds of moving to a higher level of voting engagement. This positive influence of education on voting engagement persists across all three models, with odds ratios of 1.315 (M1), 1.312 (M2), and 1.387 (M3), highlighting the consistent and statistically significant impact of education on electoral participation.

We also observe a similar pattern in our VEI models, such as age again emerges as a significant predictor of voting engagement, with older individuals more likely to exhibit higher levels of engagement, as evidenced by the odds ratios of 1.025 (M2) and 1.030 (M3). Gender shows a slight negative effect in Models 2 and 3, suggesting that females have slightly lower odds of higher voting engagement than males. The gender gap is 0.05 (M2) and 0.0495 (M3), meaning females have lower odds of higher voting engagement index than males, which is statistically significant.

Again, we observe that. Race plays a notable role in voting engagement, with Black and Middle Eastern individuals showing increased odds of higher engagement levels in Models 2 and 3. Interestingly, Native American individuals demonstrate a decreased likelihood of higher voting engagement in both models. Similarly, we observe that socioeconomic factors, such as income levels and home ownership status, are positively associated with voting engagement, indicating that higher income and home ownership are linked to higher levels of electoral participation. Political party affiliation also emerges as a crucial predictor, with Democrats, Republicans, and Independents all showing higher odds of higher voting engagement than those without party affiliation.

**Unique Contributions and Future Works:**

Our research shows that attained education consistently significantly impacts voting behavior across all four models (**see Figure 1**). Specifically, higher levels of education are associated with a clear intention to vote in upcoming elections, validated voter registration status, and, finally, a validated vote in the election. This trend persists in the composite additive index of the voting engagement index (VEI), underscoring education as a meaningful and positive reinforcement factor in engaging with the voting process within communities. Clear partisanship with the attained level of education is a strong indicator of voting engagement. Additionally, the effect of education is more pronounced in expressing a clear intention to vote (**see Figure 2**); with each attained level of education, we see *significant upward trends toward* expressing a clear intention to vote.

Interestingly, our findings reveal a nuanced pattern in gender disparity in voting behavior. Specifically, if we focus on the intention to vote, we see no significant difference between males and females in expressing a clear NO, Undecided, or maybe (**See Figure 3**). Also, we observe that increased attained education lowers both no intention to vote and vote hesitancy (undecided and maybe). The interesting observation is that an increase in schooling increases the clear intention to vote, but females' apparent intention to vote is lower than that of males (**see Figure 4**). The gender disparity in voting behavior gradually reduces when analyzed through the composite voting engagement index, indicating that increased educational attainment among women reduces this disparity in voting engagement. But we also acknowledge that gender disparity is very much present in our current society; marital status, social norms, and race and ethnicity need to be considered deeply before claiming that only education and/or education-associated and socioeconomic status are sufficient to reduce the gender gap in voting behavior in U.S. elections.

When we observe the voting behavior among Native Americans (White as reference category), we observe that the effect of educational attainment is significantly lower among Native Americans (**see Figure 5**). Although we see an upward trend among Native Americans, an increase in education is associated with a more positive intention to vote. But unfortunately, that clear and positive intention to vote does not translate equally in voter registration and voting in elections. The study's findings align with existing theories, confirming that education directly impacts civic participation (Nie et al., 1996) and supports the Partisan Mobilization theory (Dalton, 2004) and the Resource Model of Political Involvement (Brady et al., 1995, 2015; Schlozman et al., 2018), which suggest that possessing specific resources, such as education, can enhance voting engagement. We also note that higher levels of educational attainment serve as a positive reinforcement factor for Black and Hispanic Americans. However, this trend does not extend to Native Americans, who consistently exhibit lower levels of engagement across all four models, regardless of their educational attainment.

Additionally, our analysis indicates that increased income levels and homeownership significantly enhance voting behavior among respondents. Partisan identity emerges as a strong motivator in the Cooperative Congressional Election Study (CCES) survey, with respondents with a clear party affiliation being more likely to engage in all three paradigms of voting engagement. The voting engagement index further reveals that education, race, income, homeownership, and partisan identity drive holistic voting engagement behavior.

As the 2024 election approaches within six months, and our focus remains on federal-level data, we observe that women and Native American men are less likely to engage in voting behaviors, such as expressing a clear intention to vote, registering, and ultimately voting, regardless of their higher levels of educational attainment, income, homeownership, and political identity. This study recommends targeted grassroots-level campaigns and outreach to community members historically overlooked in voting behavior to mobilize the voter base and enhance civic engagement among the entire voting-age population in the United States. Furthermore, this study contributes to the understanding that attained education levels increase voting behavior participation among African and Hispanic Americans, and it advocates for specific recommendations to address these findings, but females and Native Americans' apparent intention to vote is unfortunately fully realized all the way voter registration and voting participation. We aim to explore further the nuanced factors that create these barriers among women and Native American voters in the United States and recommend more specific policies to enhance voter participation among all voting age population in the United States to ensure democratic participation for everyone eligible in future elections.

Berinsky, A. J., & Lenz, G. S. (2011). Education and Political Participation: Exploring the Causal Link. *Political Behavior*, *33*(3), 357–373.

Beyer, A., Knutsen, C. H., & Rasch, B. E. (2014). Election Campaigns, Issue Focus and Voting Intentions: Survey Experiments of Norwegian Voters. *Scandinavian Political Studies*, *37*(4), 406–427. https://doi.org/10.1111/1467-9477.12029

Bol, D., Gschwend, T., Zittel, T., & Zittlau, S. (2021). The importance of personal vote intentions for the responsiveness of legislators: A field experiment. *European Journal of Political Research*, *60*(2), 455–473. https://doi.org/10.1111/1475-6765.12408

Boyte, H. C. (2005). Reframing Democracy: Governance, Civic Agency, and Politics. *Public Administration Review*, *65*(5), 536–546. https://doi.org/10.1111/j.1540-6210.2005.00481.x

Brade, R., Piopiunik, M., & Brade, R. (2016). Education and Political Participation. *Ifo DICE Report*, *14*(01), 70–73.

Brady, H. E., Schlozman, K. L., & Verba, S. (2015). Political Mobility and Political Reproduction from Generation to Generation. *The ANNALS of the American Academy of Political and Social Science*, *657*(1), 149–173. https://doi.org/10.1177/0002716214550587

Brady, H. E., Verba, S., & Schlozman, K. L. (1995). Beyond Ses: A Resource Model of Political Participation. *The American Political Science Review*, *89*(2), 271–294. https://doi.org/10.2307/2082425

Burden, B. C. (2009). The dynamic effects of education on voter turnout. *Electoral Studies*, *28*(4), 540–549. https://doi.org/10.1016/j.electstud.2009.05.027

Afzal, Ganga, Roy, Thompson (MPSA 2024)

Franck, T. M. (1992). The Emerging Right to Democratic Governance. *American Journal of International Law*, *86*(1), 46–91. https://doi.org/10.2307/2203138

Garnett, H. A. (2022). Registration Innovation: The Impact of Online Registration and Automatic Voter Registration in the United States. *Election Law Journal: Rules, Politics, and Policy*, *21*(1), 34–45. https://doi.org/10.1089/elj.2020.0634

Geldrop, P. D. van. (2020, February 28). *What drives you to vote? : A study of the predictors of voting intention of young adults from Hengelo during local elections.* [Info:eu-repo/semantics/masterThesis]. University of Twente. https://essay.utwente.nl/80729/

Glynn, C. J., Huge, M. E., & Lunney, C. A. (2009). The Influence of Perceived Social Norms on College Students' Intention to Vote. *Political Communication*, *26*(1), 48–64. https://doi.org/10.1080/10584600802622860

Gutmann, A. (1999). *Democratic Education (Revised edition)*. Princeton University Press. https://www.jstor.org/stable/j.ctt7sdfv

Hicks, W. D., McKee, S. C., Sellers, M. D., & Smith, D. A. (2015). A Principle or a Strategy? Voter Identification Laws and Partisan Competition in the American States. *Political Research Quarterly*, *68*(1), 18–33. https://doi.org/10.1177/1065912914554039

Hyman, H. H. (1959). *Political Socialization: A Study in the Psychology of Political Behavior*. Free Press.

Jennings, M. K., Stoker, L., & Bowers, J. (2009). Politics across Generations: Family Transmission Reexamined. *The Journal of Politics*, *71*(3), 782–799. https://doi.org/10.1017/s0022381609090719Afzal, Ganga, Roy, Thompson (MPSA 2024)

**Table 1: Descriptive Statistics**

| Predictors/Outcomes | Range | SD | Min. | Max. |
|---|---|---|---|---|
| **Dependent Variables** | | | | |
| **Intention to Vote**<br>-(Self-reported turnout pre-election wave)<br>-First DV for the set of three models<br>-Ordered Logistic Regression (Odds Ratio) | 0 = No, 1 = Undecided, 2 = Probably, 3 = Yes | 0.90 | 0 | 3 |
| **Registration Status**<br>-Second DV for the set of three models<br>-Binomial Logistic Regression (Odds Ratio) | 0 = Inactive/Other, 1 = Active | 0.50 | 0 | 1 |
| **Voted Bin**<br>-Third DV for the set of three models<br>-Binomial Logistic Regression (Odds Ratio) | 0 = No, 1 = Yes | 0.49 | 0 | 1 |
| **Voting Engagement Index (VEI)**<br>-Fourth DV for the set of three models<br>-Ordered Logistic Regression (Odds Ratio) | 0 = No intention to vote, no registration, and no vote.<br>1 = Clear Intention to vote.<br>2 = Clear Intention and registration.<br>3 = Clearly defined intention, registration, and vote (verified) | 1.12 | 0 | 3 |
| **Independent Variables** | | | | |
| Education Level | 0 = No HS, 1 = HS Graduate, 2 = Some College/2-Year, 3 = 4-Year and More | 0.87 | 0 | 3 |
| Age | 18 - 109 | 16.72 | 18 | 109 |
| Gender | 0 = Male, 1 = Female | 0.50 | 0 | 1 |
| Race | 1 = White, 2 = Black, 3 = Hispanic, 4 = Asian, 5 = Native American, 6 = Mixed, 7 = Other, 8 = Middle Eastern | 1.28 | 1 | 8 |
| Income | 0 = Less than $30,000, 1 = $30,000 - $50,000, 2 = $50,000 - $70,000, 3 = $70,000 - $100,000, 4 = More than $100,000 | 1.46 | 0 | 4 |
| Home Status | 0 = Rent/Other, 1 = Own | 0.49 | 0 | 1 |
| Political Ideology | 0 = Very Conservative, 1 = Conservative, 2 = Moderate, 3 = Liberal, 4 = Very Liberal | 1.16 | 0 | 4 |
| Democrat | 1 = Yes, 0 = No | 0.48 | 0 | 1 |
| Republicans | 1 = Yes, 0 = No | 0.44 | 0 | 1 |
| Independent | 1 = Yes, 0 = No | 0.43 | 0 | 1 |
| Other & Not Sure (reference category) | 1 = Yes, 0 = No | 0.28 | 0 | 1 |
| General 2016 Participation | 0 = No, 1 = Yes | 0.31 | 0 | 1 |
| General 2012 Participation | 0 = No, 1 = Yes | 0.28 | 0 | 1 |
| General 2020 Participation (reference category) | 0 = No, 1 = Yes | 0.30 | 0 | 1 |
| Midterm 2018 Participation | 0 = No, 1 = Yes | 0.30 | 0 | 1 |
| Midterm 2014 Participation | 0 = No, 1 = Yes | 0.29 | 0 | 1 |
| Midterm 2022 Participation (reference category) | 0 = No, 1 = Yes | 0.30 | 0 | 1 |

Please note that. For the number of observations (N) in Stata, we ran a case to ensure every model had consistent complete cases. For all models in this study, we have 294,999 responses, complete cases for each dependent and independent variable.

The Voting Engagement Index is an additive index calculated as the sum of recoded intent_to_vote, reg_status_bin, and voted_bin variables, with the following recording:

intent_to_vote: 0 = 0, 1 = 0.25, 2 = 0.5, 3 = 1
reg_status_bin: 0 = 0, 1 = 1
voted_bin: 0 = 0, 1 = 1

I coded the intent to vote variable in such a manner if there are some positive intentions to vote because it captures a Likert scale; every step towards positive intention is translated with incremental 0.25 positive steps towards clearly defined intention to vote (pre-election wave). Secondly, the intention to vote variable only captures starting from the 2012 general election and captures the 2012, 2016, and 2020 general elections and the 2014, 2018, and 2022 midterm elections. We use the 2020 general election and 2022 as reference categories to capture the effects of the index on voting behavior. The LIFO approach to use 2022 mid-term as a reference category and the 2020 general election is adopted in this series of models because by utilizing the most recent elections as a reference category, we aim to uncover the nuances and specific factors that impact, influence, and affect voting behaviors.

# Table 2: Intention to Vote (DV) - Ologit Odds Ratio

| Variable | Model 1 | Model 2 | Model 3 |
|---|---|---|---|
| **Dependent Variable: Intention to Vote** <br> 0 = No, 1 = Undecided, 2 = Probably, 3 = Yes | | | |
| education_level | 1.717*** (0.015) | 1.669*** (0.017) | 1.710*** (0.017) |
| age | | 1.033*** (0.001) | 1.034*** (0.001) |
| gender | | 0.785*** (0.013) | 0.776*** (0.013) |
| race_black | | 1.590** (0.235) | 1.620** (0.250) |
| race_hispanic | | 1.545** (0.231) | 1.619** (0.252) |
| race_asian | | 1.270 (0.190) | 1.340 (0.209) |
| race_native_american | | 0.627** (0.096) | 0.677* (0.108) |
| race_other_mixed | | 1.531* (0.264) | 1.590** (0.283) |
| race_middle_eastern | | 1.718*** (0.265) | 1.789*** (0.287) |
| income | | 1.215*** (0.008) | 1.223*** (0.008) |
| home_status | | 1.435*** (0.027) | 1.399*** (0.026) |
| political_ideology | | 1.000 (0.009) | 1.011 (0.009) |
| democrat | | 3.829*** (0.126) | 3.762*** (0.124) |
| republican | | 3.093*** (0.106) | 3.171*** (0.108) |
| independent | | 1.844*** (0.059) | 1.797*** (0.057) |
| midterm2014 | | 0.839*** (0.017) | 1.224*** (0.028) |
| midterm2018 | | 0.597*** (0.012) | 0.860*** (0.020) |
| general2012 | | | 2.605*** (0.074) |
| general2016 | | | 1.800*** (0.042) |
| **Cutpoints** | | | |
| cut1 | 0.174*** (0.003) | 3.046*** (0.468) | 5.015*** (0.804) |
| cut2 | 0.331*** (0.006) | 6.115*** (0.939) | 10.160*** (1.628) |
| cut3 | 0.643*** (0.012) | 12.981*** (1.994) | 21.842*** (3.502) |
| **Fit Statistics** | | | |
| Number of obs | 294,999 | 294,999 | 294,999 |
| Log-likelihood | -113,000 | -103,000 | -102,000 |
| BIC | 225,554.62 | 206,797.13 | 204,604.13 |
| AIC | 225,512.24 | 206,585.23 | 204,371.04 |

*p < 0.05, **p < 0.01, ***p < 0.001

Please note that all the coefficients in the table are presented as odds ratios (b) with standard errors (S.E.) in parentheses. Coefficients marked with asterisks indicate statistical significance at the specified levels.

# Table 3: Registration to Vote (DV) - Logit Model

| Variable | Model 1 | Model 2 | Model 3 |
|---|---|---|---|
| **Dependent Variable: Registration to Vote** <br> 0 = Inactive/Other, 1 = Active | | | |
| education_level | 1.143*** (0.008) | 1.166*** (0.009) | 1.238*** (0.010) |
| age | | 1.019*** (0.000) | 1.024*** (0.000) |
| gender | | 1.047*** (0.012) | 1.039** (0.012) |
| race_black | | 1.591** (0.260) | 1.766*** (0.304) |
| race_hispanic | | 1.214 (0.200) | 1.384 (0.239) |
| race_asian | | 1.064 (0.176) | 1.242 (0.215) |
| race_native_american | | 0.661* (0.111) | 0.803 (0.141) |
| race_other_mixed | | 1.223 (0.216) | 1.306 (0.242) |
| race_middle_eastern | | 1.478* (0.247) | 1.619** (0.284) |
| income | | 0.994 (0.004) | 1.011* (0.005) |
| home_status | | 1.140*** (0.016) | 1.073*** (0.015) |
| political_ideology | | 0.995 (0.006) | 1.020** (0.007) |
| democrat | | 1.646*** (0.047) | 1.589*** (0.047) |
| republican | | 1.444*** (0.041) | 1.500*** (0.045) |
| independent | | 1.336*** (0.038) | 1.248*** (0.036) |
| midterm2014 | | 1.862*** (0.031) | 4.195*** (0.075) |
| midterm2018 | | 1.782*** (0.028) | 3.931*** (0.068) |
| general2012 | | | 6.127*** (0.120) |
| general2016 | | | 3.691*** (0.060) |
| _cons | 0.999 (0.015) | 0.144*** (0.024) | 0.040*** (0.007) |
| **Fit Statistics** | | | |
| Number of obs | 294,999 | 294,999 | 294,999 |
| Log-likelihood | -110,000 | -105,000 | -97,907 |
| BIC | 220,431 | 210,520 | 196,065 |
| AIC | 220,409 | 210,329 | 195,853 |

*p < 0.05, **p < 0.01, ***p < 0.001

## Table 4: Verified Voting (DV) - Logit Model

| Variable | Model 1 | Model 2 | Model 3 |
|---|---|---|---|
| **Dependent Variable: Verified Voting** | | | |
| education_level | 1.236*** (0.008) | 1.240*** (0.009) | 1.316*** (0.010) |
| age | | 1.025*** (0.000) | 1.031*** (0.000) |
| gender | | 1.012 (0.012) | 1.002 (0.012) |
| race_black | | 1.347 (0.230) | 1.471* (0.270) |
| race_hispanic | | 0.933 (0.160) | 1.038 (0.191) |
| race_asian | | 0.845 (0.146) | 0.967 (0.179) |
| race_native_american | | 0.532*** (0.093) | 0.639* (0.120) |
| race_other_mixed | | 1.082 (0.199) | 1.142 (0.224) |
| race_middle_eastern | | 1.277 (0.222) | 1.376 (0.256) |
| income | | 1.029*** (0.005) | 1.049*** (0.005) |
| home_status | | 1.212*** (0.017) | 1.150*** (0.016) |
| political_ideology | | 1.012 (0.006) | 1.037*** (0.007) |
| democrat | | 1.639*** (0.046) | 1.580*** (0.045) |
| republican | | 1.505*** (0.043) | 1.557*** (0.045) |
| independent | | 1.326*** (0.037) | 1.238*** (0.036) |
| midterm2014 | | 1.262*** (0.020) | 2.783*** (0.049) |
| midterm2018 | | 1.432*** (0.021) | 3.105*** (0.052) |
| general2012 | | | 5.482*** (0.104) |
| general2016 | | | 3.540*** (0.058) |
| _cons | 0.647*** (0.010) | 0.082*** (0.014) | 0.023*** (0.004) |
| **Fit Statistics** | | | |
| Number of obs | 294,999 | 294,999 | 294,999 |
| Log-likelihood | -111,000 | -105,000 | -98,353 |
| BIC | 221,842 | 209,609 | 196,958 |
| AIC | 221,821 | 209,418 | 196,746 |

*p < 0.05, **p < 0.01, ***p < 0.001

## Table 5: Voting Engagement Index (DV) - Ologit Model

| Variable | Model 1 | Model 2 | Model 3 |
|---|---|---|---|
| **Dependent Variable: Voting Engagement Index**<br>0 = No intention to vote, no registration, and no vote.<br>1 = Clear Intention to vote.<br>2 = Clear Intention and registration.<br>3 = Clearly defined intention, registration, and vote (verified) | | | |
| education_level | 1.315*** (0.009) | 1.312*** (0.010) | 1.387*** (0.010) |
| **age** | | **1.025*** (0.000)** | **1.030*** (0.000)** |
| **gender** | | **0.970** (0.011)** | **0.956*** (0.011)** |
| **race_black** | | **1.627** (0.248)** | **1.738*** (0.284)** |
| **race_hispanic** | | **1.260 (0.193)** | **1.401* (0.230)** |
| race_asian | | 1.092 (0.168) | 1.231 (0.203) |
| **race_native_american** | | **0.645** (0.100)** | **0.763 (0.127)** |
| race_other_mixed | | 1.297 (0.216) | 1.375 (0.243) |
| **race_middle_eastern** | | **1.574** (0.246)** | **1.689** (0.282)** |
| income | | 1.050*** (0.004) | 1.071*** (0.005) |
| home_status | | 1.252*** (0.016) | 1.193*** (0.015) |
| political_ideology | | 0.995 (0.006) | 1.018** (0.006) |
| democrat | | 2.396*** (0.072) | 2.298*** (0.069) |
| republican | | 2.062*** (0.063) | 2.115*** (0.064) |
| independent | | 1.665*** (0.050) | 1.543*** (0.046) |
| midterm2014 | | 1.396*** (0.020) | 2.847*** (0.045) |
| midterm2018 | | 1.303*** (0.021) | 2.596*** (0.045) |
| general2012 | | | 5.181*** (0.095) |
| general2016 | | | 3.204*** (0.049) |
| **Cutpoints** | | | |
| cut1 to cut9 | Various*** | Various*** | Various*** |
| **Fit Statistics** | | | |
| Number of obs | 294,999 | 294,999 | 294,999 |
| Log-likelihood | -242,000 | -233,000 | -226,000 |
| BIC | 483,426 | 466,979 | 452,692 |
| AIC | 483,320 | 466,703 | 452,395 |

*p < 0.05, **p < 0.01, ***p < 0.001

Figure 1: Effects of Education on Intention to Vote

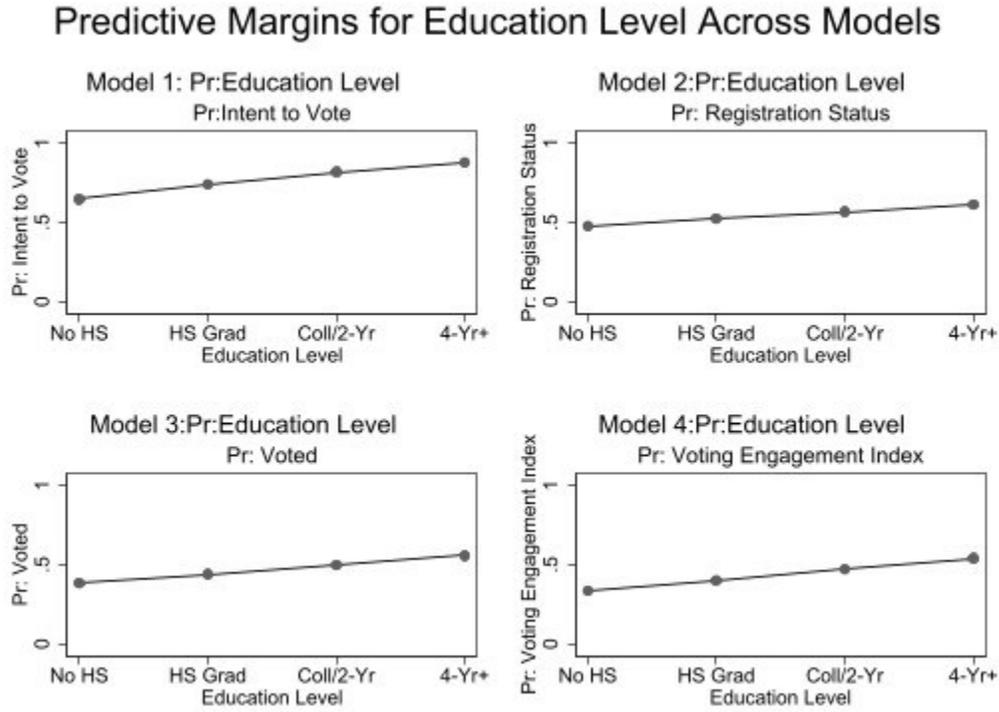

Figure 1: Predictive Margins across all four comprehensive models of how attained education impacts intention to vote, registration to vote and voting, and Voting Engagement Index.

**Figure 2:**

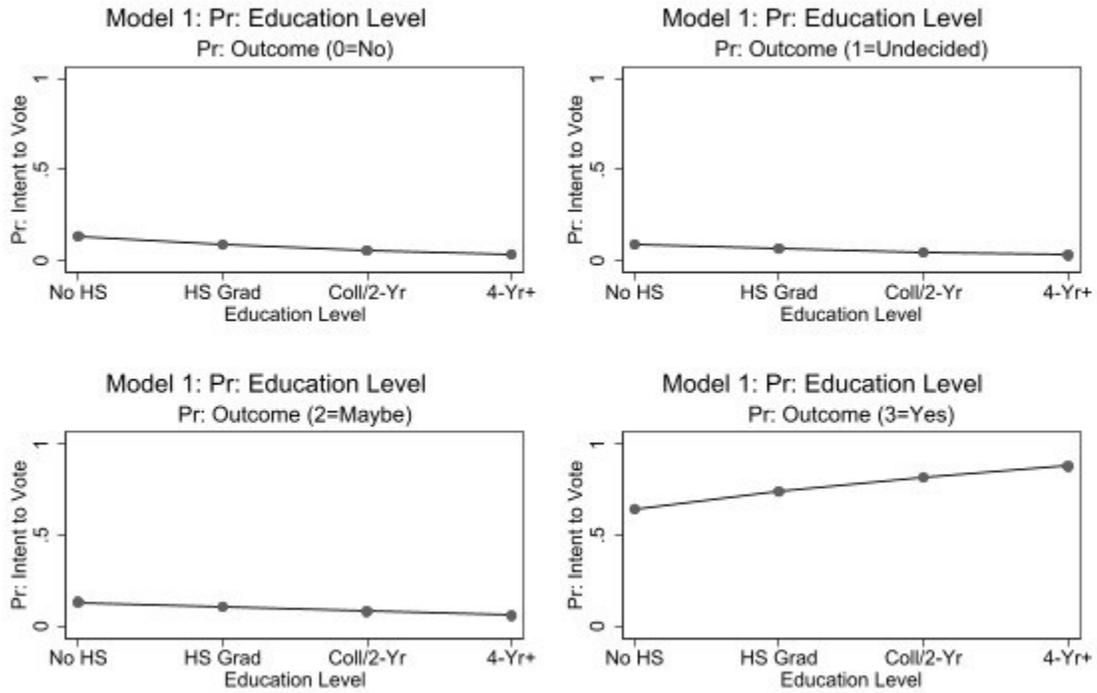

Figure 2: Predictive Margins across attained education impact intention to vote to stratify all four outcomes.

**Figure 3:**

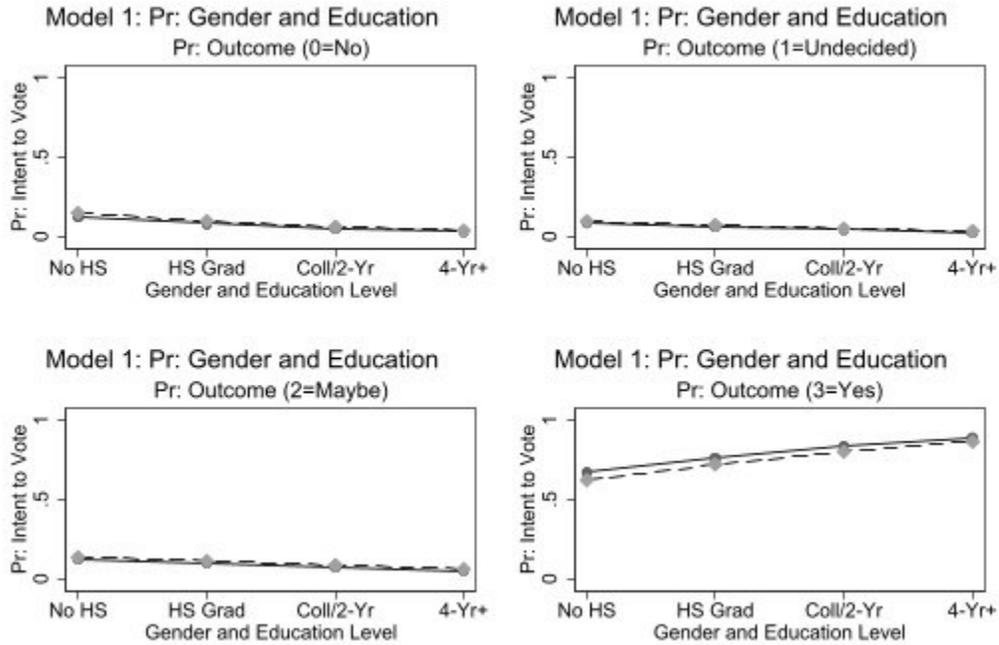

Figure 3: Effect of Education on Gender; How education and gender impact the clear intention to vote (at all levels)

**Figure 4:**

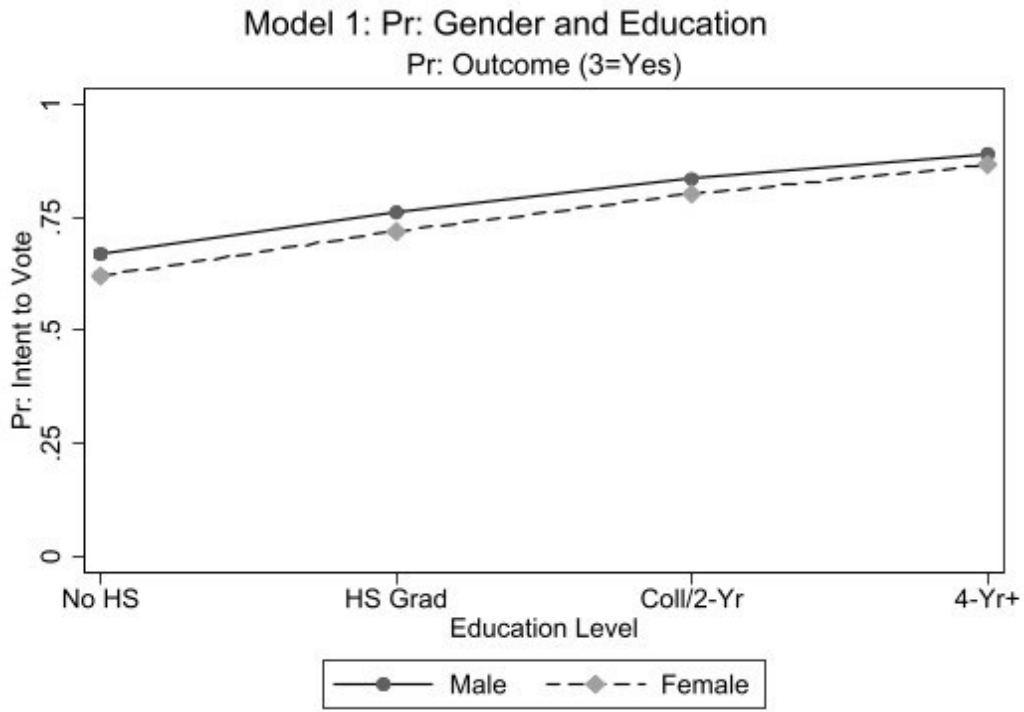

Figure 4: Effect of Education on Gender; how educated females express their clear intention to vote.

**Figure 5:**

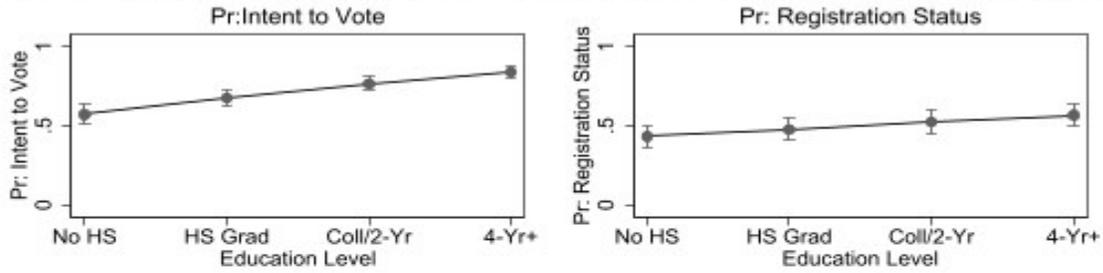

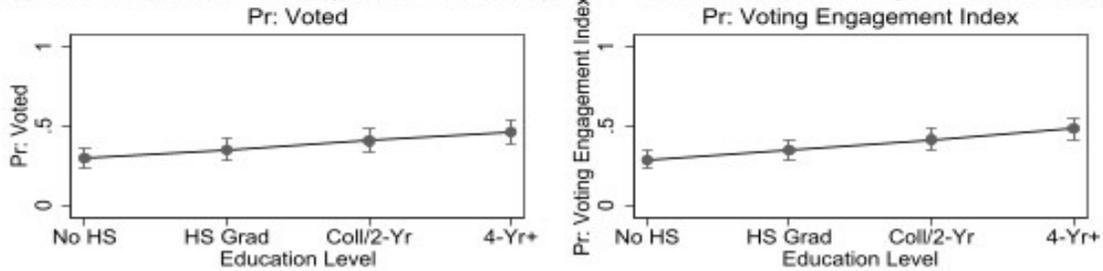